\documentclass[prd,eqsecnum, showpacs,nofootinbib]{revtex4}

\global\arraycolsep=2pt
\usepackage{stmaryrd}
\usepackage{amsmath}
\usepackage{amssymb}
\usepackage{graphicx}
\usepackage{textcomp}
\usepackage{calrsfs}
\usepackage{yfonts}
\usepackage{bm}

\newcommand{\ben}{\begin{equation*}}
\newcommand{\een}{\end{equation*}}
\newcommand{\bean}{\begin{eqnarray*}}
\newcommand{\eean}{\end{eqnarray*}}
\newcommand{\bnabla}{\mbox{\boldmath{$\nabla$}}}

\newcommand{\nn}{\nonumber}
\newcommand{\be}{\begin{equation}}
\newcommand{\ee}{\end{equation}}
\newcommand{\bea}{\begin{eqnarray}}
\newcommand{\eea}{\end{eqnarray}}

\DeclareMathOperator{\Tr}{Tr}
\begin{document}

\title{How does Casimir energy fall?  IV.\\Gravitational interaction of 
regularized quantum vacuum energy}

\author{K. A. Milton}

\affiliation{Laboratoire Kastler Brossel, Universit\'e Pierre et Marie Curie,
4, Place Jussieu Case 74, F-75252 Paris Cedex 05,  France}
\altaffiliation{Permanent address:
H. L. Dodge Department of Physics and Astronomy,
University of Oklahoma, Norman, OK 73019 USA}
\email{milton@nhn.ou.edu}

\author{K. V. Shajesh}
\affiliation{Department of Physics, Southern Illinois University--Carbondale,
Carbondale, IL 62901 USA}
\author{S. A. Fulling}
\affiliation{Departments of Mathematics and Physics, Texas A\&M University,
College Station, TX 77843 USA}
\author{Prachi Parashar}
\affiliation{H. L. Dodge Department of Physics and Astronomy,
University of Oklahoma, Norman, OK 73019 USA}
\date{\today}

\begin{abstract}
Several years ago we demonstrated that the Casimir energy for perfectly
reflecting  and imperfectly reflecting
 parallel plates gravitated normally, that is, obeyed the
equivalence principle.  At that time the divergences in the theory were treated
only formally, without proper regularization, and the coupling to gravity
was limited to  the canonical energy-momentum-stress tensor.  
Here we strengthen the result by removing both of those 
limitations. We consider, as a toy model, massless scalar fields interacting
with semitransparent ($\delta$-function) potentials defining parallel plates,
which become Dirichlet plates for strong coupling.
 We insert space and time point-split regulation parameters,
and obtain well-defined contributions to the self-energy of each plate, and
the interaction energy between the plates.  (This self-energy does not
vanish even in the conformally-coupled, strong-coupled limit.)
We also compute the local energy density, which requires regularization near
the plates. In general, the energy density includes a surface energy
 that resides precisely
on the boundaries.  This energy is also regulated.  The gravitational
interaction of this well-defined system is then investigated, and it
is verified that the equivalence principle is satisfied.

\end{abstract}


\maketitle
\section{Introduction}
The subject of quantum vacuum energy (the Casimir effect) dates from the
same year as the discovery of renormalized quantum electrodynamics, 1948
\cite{casimir}.
It puts the lie to the  presumption that  zero-point energy is 
 not observable \cite{nernst,pauli,kragh}. 
On the other hand, because of the severe divergence structure
of the theory, controversy has surrounded it from the beginning.
Here we will deal with divergences carefully, by using point-splitting
in space and time.

The volume divergence, sometimes called the bulk term, is rather easily
isolated, and apparently has no physical consequences, since it does
not refer to anything but the properties of empty space.  Once bodies
are introduced, additional divergences appear.
Sharp boundaries, and even soft ones, 
give rise to divergences in the local energy density near
the surface \cite{Deutsch:1978sc,Milton:2004vy,Milton:2011iy}.
Curvature introduces additional divergences,
and if the surfaces possess discontinuities such as corners, there
will be additional divergent terms associated with these. 
These divergences  may make it impossible to extract meaningful self-energies
of single objects, the cancellations for the electromagnetic field at
perfectly conducting planes \cite{casimir} and spheres \cite{Boyer:1968uf}
being accidental \cite{Deutsch:1978sc}.  How can
something finite be meaningfully extracted from this wealth of infinities
(which are actually finite, but very large, if a physically reasonable
microscopic cutoff is inserted)?
These  objections have been most forcefully presented by 
Graham, Jaffe, et al.\ \cite{Graham:2003ib,Graham:2002fw}, and by Barton
\cite{barton,barton2},
but they date back to 
Deutsch and Candelas \cite{Deutsch:1978sc}.

In fact, it has appeared for some time
 that these surface divergences can be dealt with successfully
in a process of renormalization  (see for example,
Refs.~\cite{Milton:2013yqa,Milton:2013xia})
and that finite self-energies, in a generalization of the sense
of Boyer \cite{Boyer:1968uf}, may be extracted.  
So in this paper we will consider not only the
universally recognized unambiguous Casimir interaction energies, but also the
divergent, but regulated,  self-energies of the separate bodies, 
here planar objects. It is critical to do this, because
 gravity couples to the local energy-momentum tensor, and such
surface divergences and self-energies  promise serious difficulties.
 How is the completely finite Casimir interaction 
energy of a pair of parallel conducting plates, as well as the divergent
self-energies of non-ideal plates,
accelerated by gravity? We must also address the issue 
 of the renormalization of Einstein's equations
resulting from singular Casimir surface energy densities 
\cite{Estrada:2008zza,Estrada:2012yn}.
The resolution of these questions
 turns out to be surprisingly less straightforward
than the reader might suspect! 

 In the remainder of the introduction we shall recapitulate the previous 
papers in this series \cite{Fulling:2007xa,Milton:2007ar,Shajesh:2007sc}.
We use natural units (in particular, $c=1$), so that energy is 
identified with mass, and acceleration has the units of inverse length.

\subsection{Gravitational coupling to an ideal Casimir apparatus}
Brown and Maclay \cite{brown}
showed that, for parallel  perfectly
conducting plates separated by a distance $a$ in the $z$-direction, 
the electromagnetic stress
tensor acquires the vacuum expectation value between the plates
\be \langle T^{\mu\nu}\rangle
=\frac{\mathcal{E}_C}a \mbox{diag} (1, -1, -1, 3),
\quad {\mathcal{E}_C=-\frac{\pi^2}{720 a^3},}\ee
$\mathcal{E}_C$ being Casimir's energy per unit area. 
Outside the plates the value of $\langle T^{\mu\nu}\rangle$ is $0$.  
What is  the gravitational interaction of this Casimir apparatus?
As shown in Ref.~\cite{Fulling:2007xa},
 this question can be most simply addressed through
use of the gravitational definition of the energy-momentum tensor,
\be
\delta W_m\equiv -\frac12\!\int(dx) \sqrt{-g}\,\delta g^{\mu\nu}T_{\mu\nu}
=\frac12\!\int(dx) \sqrt{-g}\,\delta g_{\mu\nu}T^{\mu\nu}.\label{var}
\ee
For a weak field,
\be
g_{\mu\nu}=\eta_{\mu\nu}+2h_{\mu\nu},
\ee
so if we think of turning
on the gravitational field as a small perturbation, we can ignore $\sqrt{-g}$.
The gravitational energy, for a static situation, is therefore given by
($\delta W=-\int dt\,\delta E$)
\be
E_g=-\int (d\mathbf{x}) h_{\mu\nu}T^{\mu\nu}.\label{ge}
\ee

\begin{figure}
\centering
\includegraphics{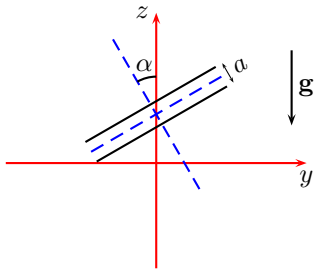}
\caption{\label{gfig}
A Casimir apparatus of two parallel plates, the normal to which makes
an angle $\alpha$ with respect to the direction of gravity, the negative $z$
axis. The parallel plates are indicated by the heavy lines.}
\label{fig1}
\end{figure}

The Fermi metric locally describes an inertial coordinate system:
\be
h_{00}=-gz,\quad h_{0i}=h_{ij}=0,
\ee
which is appropriate for describing a constant gravitational
field.  Let us consider a Casimir apparatus of parallel plates
separated by a distance $a$, with transverse dimensions $L\gg a$.
Let the apparatus be oriented at an angle $\alpha$ with respect to the
direction of gravity.  
The Cartesian coordinate
system attached to the earth is $(x, y, z)$, where
$z$ is the direction of $-\mathbf{g}$.  
See Fig.~\ref{gfig}.
Now we calculate the gravitational energy
\be
E_g=\int (d\mathbf{x})gzT^{00}
=g\mathcal{E}_CL^2 z_0+K,
\ee
where $K$ is a constant, independent of the center $z_0$ of the
apparatus.  Thus, the gravitational force per area $A=L^2$ 
on the apparatus is independent of orientation:
\be
\mathcal{F}\equiv
\frac{F}A=-\frac{\partial E_g}{A\partial z_0}=-g\mathcal{E}_C,
\label{f0}
\ee
a small upward push.  
Therefore, $\mathcal{E}_C$ just adds to the mass energy of the plates,
precisely in accordance with the equivalence principle.

\subsection{Uniform acceleration, semitransparent plates}

A more exact relativistic
 calculation is based on the use of Rindler coordinates to
describe constant acceleration \cite{Milton:2007ar}.  
In the balance
of this paper, for simplicity, we will consider scalar fields interacting
with $\delta$-function (semitransparent) plates.
Relativistically, uniform (but necessarily $\xi$ dependent)
 acceleration is described by hyperbolic motion,
\be
t=\xi\sinh\tau,\quad z=\xi\cosh\tau,
\ee
which induces the metric
\be
dt^2-dz^2-d\mathbf{r}_\perp^2=\xi^2d \tau^2-d\xi^2-d\mathbf{r}_\perp^2.
\ee
The d'Alembertian operator has cylindrical form
\be
\partial^2=-\left(\frac\partial{\partial t}\right)^2
+\left(\frac\partial{\partial z}\right)^2+\nabla^2_\perp
=-\frac1{\xi^2}\left(\frac\partial{\partial\tau}\right)^2+\frac1\xi
\frac\partial{\partial\xi}\left(\xi\frac\partial{\partial\xi}\right)
+\nabla^2_\perp.
\ee

For two   semitransparent ($\delta$-function) 
\cite{Scandurra:2000qz,Milton:2004vy}
plates at $\xi_1$ and $\xi_2$, the Green's function can be written as
\be
G(x,x')=\int\frac{d\omega}{2\pi}\frac{(d\mathbf{ k_\perp})}{(2\pi)^2}
e^{-i\omega(\tau-\tau')}e^{i\mathbf{k_\perp\cdot(r-r')_\perp}}g(\xi,\xi'),
\ee
where the reduced Green's function satisfies
\be
\left[-\frac{\omega^2}{\xi^2}-\frac1\xi\frac\partial{\partial\xi}\left(\xi\frac
\partial{\partial\xi}\right)+k_\perp^2+\lambda_1\delta(\xi-\xi_1)
+\lambda_2\delta(\xi-\xi_2)\right]g(\xi,\xi')=
\frac1\xi\delta(\xi-\xi'),\label{rgfe}
\ee
which we recognize as just the problem of two concentric
semitransparent cylinders  \cite{CaveroPelaez:2006rt}
with the replacements  $m\to \zeta=-i\omega$ and
$\kappa\to k$. 
The explicit solution for the reduced Green's function
$g$ is given in Ref.~\cite{Milton:2007ar} in terms of modified
Bessel functions, $I_\zeta(k_\perp\xi)$, $K_\zeta(k_\perp\xi)$.

The {\it canonical\/} energy-momentum tensor for a scalar field is given by
\be
T_{\mu\nu}=\partial_\mu\phi\partial_\nu\phi
+g_{\mu\nu}\mathcal{L},\quad \mathcal{L}=-\frac12
\partial_\lambda\phi\partial^\lambda\phi-\frac12V\phi^2,
\ee
where the Lagrange density includes the $\delta$-function potential,
\be
V=\lambda_1\delta(\xi-\xi_1)+\lambda_2\delta(\xi-\xi_2).\label{dfpot}
\ee
Using the equation of motion,
\be
(-\partial^2+V)\phi=0,\label{eom}
\ee
we find the energy density to be
\be
T_{00}=\frac12\left(\frac{\partial\phi}{\partial\tau}\right)^2-\frac12\phi
\frac{\partial^2}{\partial\tau^2}\phi+\frac\xi 2\frac\partial{\partial\xi}
\left(\phi\xi\frac\partial{\partial\xi}\phi\right)+\frac{\xi^2}2\bnabla_\perp
\cdot(\phi\bnabla_\perp\phi).\label{st00}\ee
We obtain the vacuum expectation value of the stress tensor,
$\langle T_{\mu\nu}\rangle$, from the replacement
\be\langle \phi(x)\phi(y)\rangle=\frac1iG(x,y).\label{qmr}\ee

The (gravitational) force density is given by \cite{moller}
\be
f_\lambda=-\frac1{\sqrt{-g}}\partial_\nu(\sqrt{-g}T^\nu{}_\lambda)
+\frac12T^{\mu\nu}\partial_\lambda g_{\mu\nu},\label{forced}
\ee
so the gravitational force per unit area  on the system is, upon integration by
parts, %(but see Sec.~\ref{secfall}, below)
\be
\frac1g\mathcal{F}=\int d\xi \xi f_\xi=-\int\frac{d\xi}{\xi^2}T_{00}
=\int d\xi \xi\int\frac{d\hat\zeta \,(d\mathbf{k_\perp})}{(2\pi)^3}\hat\zeta^2 
g(\xi,\xi),\quad(\zeta=\hat\zeta\xi).
\ee
This is the change of momentum per unit Rindler coordinate time 
$\tau$,
which when multiplied by the gravitational acceleration at $\xi_0$,
namely, $g=1/\xi_0$, is the
gravitational force/area $\mathcal{F}$ on the Casimir energy
in an apparatus centered at Rindler position $\xi_0$.
 The reader is referred to Ref.~\cite{Milton:2007ar} for details.
For the purposes here,
all we need is  the weak acceleration limit. 
This is the limit in which $\xi$, $\xi'$, $\xi_1$, and $\xi_2$ all tend to
infinity, but expanded about $\xi_0$ so that
differences such as $\xi-\xi'$ are finite.  Likewise, we rescale
$\zeta=\xi_0\hat\zeta$, and regard $\hat\zeta$ and $\kappa^2=k_\perp^2+\hat
\zeta^2$ as finite.  Then  the Green's function
reduces to exactly the expected result, for example,
between the plates,  $\xi_1<\xi,\xi'<\xi_2$
($a=\xi_2-\xi_1$)
\be
\xi_0g(\xi,\xi')\to\frac1{2\kappa}e^{-\kappa|\xi-\xi'|}
+\frac1{2\kappa\Delta}\bigg[\frac{\lambda_1\lambda_2}{(2\kappa)^2}
2\cosh\kappa(\xi-\xi')
-\frac{\lambda_1}{2\kappa}\left(1+\frac{\lambda_2}{2\kappa}
\right)e^{-\kappa(\xi+\xi'-2\xi_2)}
-\frac{\lambda_2}{2\kappa}\left(1+\frac{\lambda_1}{2\kappa}\right)
e^{\kappa(\xi+\xi'-2\xi_1)}\bigg],\label{2pgf}
\ee
where the denominator (which has a simple interpretation in terms of
multiple reflections) is
\be
\Delta=\left(1+\frac{\lambda_1}{2\kappa}\right)
\left(1+\frac{\lambda_2}{2\kappa}\right)e^{2\kappa a}
-\frac{\lambda_1\lambda_2}{4\kappa^2}.\label{Delta}
\ee
 The flat space limit also holds outside the plates.

From this follows the explicit force per unit area on the 2-plate apparatus as
\bea
\mathcal{F}&=&\frac{g}{96\pi^2 a^3}\int_0^\infty dy\,y^3\frac{1+
\frac1{y+\lambda_1a}
+\frac1{y+\lambda_2a}}{\left(\frac{y}{\lambda_1a}+1\right)
\left(\frac{y}{\lambda_2a}+1\right)e^y-1}
-\frac{g}{96\pi^2 a^3}\int_0^\infty dy\,y^2\left[\frac1{\frac{y}
{\lambda_1a}
+1}+\frac1{\frac{y}{\lambda_2a}+1}\right]\nn\\
&=&-g(\mathcal{E}_C+\mathcal{E}_{S_1}+\mathcal{E}_{S_2}),\label{ef}\eea
which is just $-g$ times the Casimir energy/area of the two semitransparent
plates, {\it including the divergent parts associated with each
plate.}  Note that the divergent parts are independent of the separation
between the plates.
The divergent terms are self-energies $\mathcal{E}_{S_{1,2}}$ which
simply renormalize the mass/area of each plate:
\be
\mathcal{E}_{\rm total}=m_1+m_2+\mathcal{E}_{S_1}+\mathcal{E}_{S_2}
+\mathcal{E}_C=M_1+M_2+\mathcal{E}_C,
\ee
and thus the gravitational force on the entire apparatus obeys the
equivalence principle
\be
\mathcal{F}=-g(M_1+M_2+\mathcal{E}_C).
\ee

This 
calculation has been implicitly carried out in the vacuum state of the 
field quantized in the Rindler coordinate system.  For completeness one 
should also consider the presence of Unruh radiation (or Hawking-Hartle 
radiation, in the case of a Schwarzschild gravitational source).  This 
complication is left for later investigation.

\begin{figure}
\centering
\includegraphics{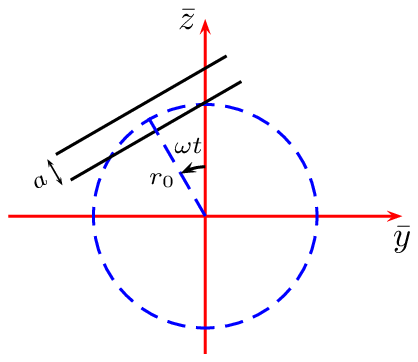}
 \includegraphics{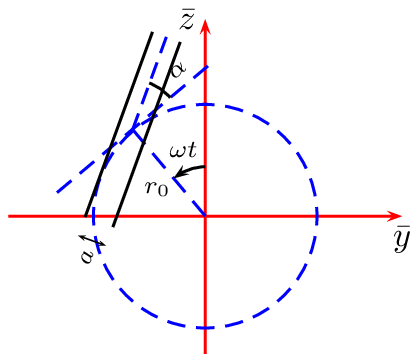}
\vspace{1.5in}
\caption{\label{fig2}
Casimir apparatus undergoing circular motion.  The Casimir
energy contributes in the usual manner to the inertial mass of the system, and
the divergent contributions to the energy renormalize the masses of
the two Casimir plates. The first panel shows the normal of the apparatus
in the radial direction, the second with that axis making an arbitrary
angle $\alpha$ with respect to the radius.}
\end{figure}

A third paper in this series \cite{Shajesh:2007sc} 
considered a
Casimir apparatus undergoing centripetal acceleration 
as shown in Fig.~\ref{fig2}.  The centripetal force on the
apparatus rotating with angular speed $\omega$, $\omega r\ll1$, is 
\be
\mathbf{F}=-\omega^2\int (d\mathbf{r})\,\mathbf{r}\,T_{00}(\mathbf{r})
=-\omega^2\mathbf{r}_{\rm CM}(m_1+m_2+E_{S_1}+E_{S_2}+E_C),\ee
where $\mathbf{r}_{\rm CM}$ is the position vector of the center of energy.
Again, the self-energies correctly renormalize the mass of the plates.

Other work demonstrating that Casimir energy possesses the correct
Einstein inertia includes Ref.~\cite{jaekel}.

%%%%NEW MATERIAL
\section{Regulated calculation of Casimir energy of parallel semitransparent
plates}
\subsection{Fundamental formulas}

In this and the following two sections we will consider Minkowski spacetime.
We will also be freely using the same symbols to represent operators
and functions, illustrated by the Green's function
\be
G(x,x')=\langle x|G|x'\rangle.
\ee
The imaginary frequency is represented by $\zeta$.

The fundamental formula for the Casimir energy can be taken to be
the famous trace-log formula,\footnote{A convincing argument for using
this as a starting point is that then the correct free energy emerges upon
replacing the imaginary frequency integral by the sum over  Matsubara 
frequencies.}
\be
E=-\frac12\int_{-\infty}^\infty \frac{d\zeta}{2\pi}\Tr\ln\mathcal{G}
.\label{fund}
\ee
From this, by formal integration by parts, one obtains another commonly-used
form
\be
E=-\int_{-\infty}^\infty \frac{d\zeta}{2\pi}\zeta^2\Tr \mathcal{G}.\label{sec}
\ee
But one might rightly be suspicious of this because the integrals are not
well-defined.  We will properly define the regulated versions of these
integrals in the following subsection.

The analysis sketched in the introduction
 may be equally well  criticized for not dealing with divergences
properly, and including manipulations with divergent integrals. 
In this paper, we will remedy this situation.  
We also wish to include arbitrary values of the conformal coupling
parameter, because these correspond to more general couplings to
gravity, and include the conformally coupled case which may have
special virtues \cite{ccj}.  We will consider two
semitransparent plates, interacting with a massless scalar field, with
the potential
\be V=\lambda_1\delta(z)+\lambda_2\delta(z-a).\label{fspot}
\ee
The time-Fourier transformed  Green's function has the form
\be
\mathcal{G}(\mathbf{r,r'};\omega)
=\int\frac{(d\mathbf{k_\perp})}{(2\pi)^2}e^{i\mathbf{
k_\perp\cdot(r-r')_\perp}}g(z,z'),\label{tftgf}\ee
where, between the plates, $0<z,z'<a$, the reduced Green's function
has precisely the form given in Eq.~(\ref{2pgf}) with $\xi$ and $\xi'$
replaced by $z$ and $z'$, $\xi_1=0$, $\xi_2=a$, and $\hat\zeta\to\zeta$.
In particular, $\kappa=\sqrt{k_\perp^2+\zeta^2}$.
In the region outside the plates, the reduced Green's function has the form
\begin{subequations}
\label{unregus}
\bea
z,z'<0:\quad g(z,z')&=&\frac{e^{-\kappa|z-z'|}}{2\kappa}
-\frac{e^{\kappa(z+z')}}{2\kappa\Delta}\left[\frac{\lambda_2}{2\kappa}\left(
1-\frac{\lambda_1}{2\kappa}\right)+\frac{\lambda_1}{2\kappa}\left(1+
\frac{\lambda_2}{2\kappa}\right)e^{2\kappa a}\right],\label{geebelow}\\
z,z'>a:\quad g(z,z')&=&\frac{e^{-\kappa|z-z'|}}{2\kappa}
-\frac{e^{\kappa(2a-z-z')}}{2\kappa\Delta}\left[\frac{\lambda_1}{2\kappa}\left(
1-\frac{\lambda_2}{2\kappa}\right)+\frac{\lambda_2}{2\kappa}\left(1+
\frac{\lambda_1}{2\kappa}\right)e^{2\kappa a}\right],
\eea
\end{subequations}
where $\Delta$ is given by Eq.~(\ref{Delta}).

\subsection{Point-split regularization}

To define the integrals, we adopt point splitting in the time and the 
transverse directions (but not in the $z$ direction, so as not
to complicate the boundary conditions):
\be \tau=t_E-t_E'\to 0,\quad {\bf R_\perp=(r-r')_\perp}\to 0.\ee
Here we have made a Euclidean rotation,
\be
\omega\to i\zeta,\quad t\to it_E,\quad \mbox{so}\quad \omega t\to -\zeta t_E.
\ee
Effectively $\mathcal{G}$ is now the cylinder kernel in the sense of
Refs.~\cite{Fulling:2003zx,Estrada:2008zza,Estrada:2012yn}.
For our transversely translationally invariant system, if we insert
Eq.~(\ref{tftgf}) into our fundamental form for the energy (\ref{fund}), 
and use the above regulator factors,  we obtain for the energy per unit
area
\be
\mathcal{E}=-\frac12\int\frac{d\zeta}{2\pi}\int\frac{(d \mathbf{k_\perp})}
{(2\pi)^2}e^{i\bm{\kappa\cdot\delta}}\Tr\ln g,\label{efund}\ee
in terms of the reduced Green's function.
Here  we have united frequency and transverse momentum as 
$\bm{\kappa}=(\zeta,\mathbf{k_\perp})$, and
similarly united the time and transverse spatial splittings as 
$\bm{\delta}=(\tau,\mathbf{\bf R_\perp})$.  Let
$\gamma$ be the angle between $\bm{\delta}$ and the time axis.  Thus,
$\gamma=0$ corresponds to time-splitting regularization, $\gamma=\pi/2$ to
transverse space-splitting.  The latter splitting is in the 
neutral direction, as defined in Ref.~\cite{Estrada:2012yn},
that is, not involved in the definition of the relevant stress-tensor
component, nor in the geometrically relevant direction.  In that
case, the integration by parts in passing to the regulated form of
the energy (\ref{sec}) is legitimate,
because the cutoff function does not depend on $\zeta$.  
In general, when integrating over
the spherical angles for $\bm{\kappa}$, $\alpha$ and $\beta$, we
encounter 
\be
f(\gamma)=\int_{-1}^1 d\cos\alpha\int_0^{2\pi}d\beta\cos^2\alpha\, 
e^{i\bm{\kappa}\cdot \bm{\delta}}\to \frac{4\pi}3,
\quad
w(\gamma)=\int_{-1}^1 d\cos\alpha\int_0^{2\pi}d\beta\, e^{i\bm{\kappa}
\cdot \bm{\delta}}\to {4\pi}.\label{cutoffs}
\ee
The limits are as $|\bm{\delta}|=\delta\to0$.  
For transverse space-splitting, the explicit forms of the cutoff functions are
\be
f(\pi/2)=4 \pi\left(-\frac{\cos\kappa\delta}{(\kappa\delta)^2}+\frac{\sin\kappa
\delta}{(\kappa\delta)^3}\right), \quad w(\pi/2)=4\pi\frac{\sin\kappa\delta}
{\kappa\delta},\label{transco}\ee
and as a result
\be
\mathcal{E}(\gamma=\pi/2)
=-\frac1{(2\pi)^3}\int_0^\infty d\kappa\,\kappa^4 f(\pi/2)\Tr g.
\label{regenpi2}
\ee
For time-splitting, the corresponding forms of the cutoff functions are
obtained from
\be
f(0)=\frac{d}{d\delta}\left[\delta f(\pi/2)\right],
\quad w(0)=w(\pi/2).\label{0topi}\ee
However, in this case 
the integration by parts leading to the regulated form of Eq.~(\ref{sec})
 proceeds as follows:
\bea
\mathcal{E}(\gamma=0)&=&-\frac12\int\frac{d\zeta}{2\pi}
\frac{(d\mathbf{k_\perp})}{(2\pi)^2}e^{i\zeta\tau}
\Tr\ln g=-\int_0^\infty \frac{d\kappa\,\kappa^2}
{(2\pi)^3}\int_{-1}^1 d\cos\alpha\int_0^{2\pi}d\beta\,\kappa\cos\alpha
\frac1{i\delta}\left(e^{i\kappa\delta\cos\alpha}-1\right)\Tr g\label{ibp},
\eea
which uses the indefinite integral
\be
\int d\zeta e^{i\zeta\tau}=\frac1{i\tau}\left(e^{i\zeta\tau}-1\right),
\ee
and, in view of the realization of $g^{-1}$ as a differential operator,
\be
g^{-1}=\zeta^2+k_\perp^2-\frac{d^2}{dz^2}+V,
\ee
we have
\be
g^{-1}\frac\partial{\partial\zeta}g=-\frac{\partial}{\partial\zeta}g^{-1} g=
-2\zeta g.
\ee
Now the integral over the angles in Eq.~(\ref{ibp}) is
\be
\int_{-1}^1 d\cos\alpha\int_0^{2\pi}d\beta\,\kappa\cos\alpha
\frac1{i\delta}\left(e^{i\kappa\delta\cos\alpha}-1\right)=\kappa^2f(\pi/2),
\ee
that is, there is no difference between time and space splitting!
The pressure anomaly \cite{Estrada:2012yn} apparently affects only 
Eq.~(\ref{sec}), not Eq.~(\ref{fund}).
Thus,
\be
\mathcal{E}(\gamma=0)=\mathcal{E}(\gamma=\pi/2)
=-\frac1{(2\pi)^3}\int_0^\infty d\kappa\,\kappa^4 f(\pi/2)\Tr g.
\label{regene}
\ee
(We have not examined other values of $\gamma$.)

Hence, if we insert the reduced Green's function given above
in Eqs.~(\ref{2pgf}) and (\ref{unregus}), we find
  the energy/area to be given by ($L_z$ is the extent
of the system in the $z$ direction)
\be
\mathcal{E}=-\int_0^\infty\frac{d\kappa\,\kappa^2}{(2\pi)^3}\kappa^2
f(\pi/2)\bigg\{
\frac{L_z}{2\kappa}+\frac1{4\kappa^2\Delta}\bigg[4(\kappa a+1)\frac{
\lambda_1\lambda_2}{(2\kappa)^2}
-2e^{2\kappa a}\left(\frac{\lambda_1+\lambda_2}{2\kappa}+2\frac{\lambda_1
\lambda_2}{(2\kappa)^2}\right)\bigg]\bigg\}.\label{gwithps}\ee

First we look at the Weyl, or bulk, term, that would be present with no 
boundaries,  corresponding to the term in Eq.~(\ref{gwithps}) 
proportional to $L_z$:
\be
E_W(\gamma=\pi/2)=-\frac{V}{8\pi^3}\int_0^\infty d\kappa \,\kappa^2f(\pi/2)
\frac12\kappa=-\frac{V}{2\pi^2\delta^4},\label{bulk}
\ee
just as expected.  If we had replaced $f(\pi/2)$ by $f(0)$ to obtain
the  corresponding time-split divergence we would have obtained
using Eq.~(\ref{0topi})
\be
E_W(\gamma=0)=\frac3{2\pi^2}\frac{V}{\delta^4},\ee
as is familiar; but as we have seen, if we regard Eq.~(\ref{fund}) rather
than Eq.~(\ref{sec}) as fundamental, this is not legitimate, and 
Eq.~(\ref{bulk}) is the bulk energy for either type of regularization.

\subsection{Self and interaction energies}
It is then straightforward to calculate the balance of the energy/area
($y=2\kappa a$):
 \be
\mathcal{E}-\mathcal{E}_W=\frac1{128\pi^3 a^3}\int_0^\infty 
\!\!\!dy\,y^2 f(\pi/2)
\left(\frac1{\frac{y}{\lambda_1 a}+1}+\frac1{\frac{y}{\lambda_2a}+1}\right)
-\frac1{96\pi^2 a^3}\int_0^\infty dy\,y^3
\frac{1+\frac1{y+\lambda_1 a}
+\frac1{y+\lambda_2 a}}{\left(\frac{y}{\lambda_1 a}+1\right)\left(
\frac{y}{\lambda_2 a}+1\right)e^y-1}.\label{regen}\ee
We have set the cutoff to zero in the second, finite term.  That term is
the same as given in Eq.~(\ref{ef}) 
 for the Casimir interaction energy $\mathcal{E}_C$ 
of parallel semitransparent plates.

The divergent term, which agrees with that in Eq.~(\ref{ef}), 
$\mathcal{E}_{S_{1,2}}$, 
when the formal replacement in Eq.~(\ref{cutoffs}) is made,
 is the sum of contributions from each plate separately, 
which are unaware of the other plate.  The self-energy of a single plate is,
as $\delta\to0$, 
\be
\mathcal{E}_{S_i}=\frac{\lambda_i}{8\pi^2}\left[\frac1{\delta^2}
-\frac{\lambda_i}8
\frac\pi\delta-\frac{\lambda_i^2}{12}\left(\ln\lambda_i\delta/2+\gamma-
\frac43\right)\right]+O(\delta), \quad i=1,2.\label{sefinl}
\ee
This is for finite $\lambda_i$, $\lambda_i\delta\ll1$. (Here $\gamma$ is
Euler's constant.)
This expansion can be found  from the heat kernel expansion given,
for example, in Ref.~\cite{Bordag:1998vs};  to compare
with our spatial-splitting results, we convert the heat kernel expansion
to the cylinder kernel expansion using the formulas in 
Ref.~\cite{Fulling:2003zx}. See also 
Refs.~\cite{Milton:2013yqa,Milton:2013xia}.
In the Dirichlet limit,
$\lambda_i\to\infty$, the self-energy is more divergent:
\be
\mathcal{E}_{S_i}=\frac1{8\pi}\frac1{\delta^3}.\label{sed}\ee
This also
corresponds to the known surface  term in the heat kernel expansion.
The total energy thus has four components:
\be
\mathcal{E}=\mathcal{E}_W+\mathcal{E}_{S_1}+\mathcal{E}_{S_2}+\mathcal{E}_{C}.
\ee
The interpretation of this result is straightforward: the Weyl term, 
$\mathcal{E}_W$, is the unobservable vacuum energy of empty space, the 
self-energies, $\mathcal{E}_{S_{1,2}}$, renormalize masses of the plates,
and only the interaction term, $\mathcal{E}_C$, is the observable Casimir
energy.

\section{Local energy density}

\subsection{Forms of stress tensor}
To answer the question of how Casimir energy interacts with gravity, we
must look at local quantities.  The stress tensor, now including the conformal
term,  for a massless scalar field is
\be
T^{\mu\nu}=\partial^\mu\phi\partial^\nu\phi-\frac12 g^{\mu\nu}(\partial_\lambda
\phi\partial^\lambda\phi+V\phi^2)
-\eta(\partial^\mu\partial^\nu-g^{\mu\nu}\partial^2)
\phi^2,\label{t00def}\ee
where $\eta$ is the conformal parameter; $\eta=1/6$ is the choice that
makes conformal invariance manifest.
Then, the Fourier-transformed expectation value of the stress tensor, 
$t^{\mu\nu}$, given by
\be
\langle T^{\mu\nu}\rangle =\int\frac{d\zeta}{2\pi}\frac{(d\mathbf{k_\perp})}
{(2\pi)^2}e^{i\zeta\tau}e^{i\mathbf{k_\perp\cdot R_\perp}}
t^{\mu\nu}(z,z')\bigg|_{
z'\to z},\label{ftt}\ee
is obtained with the quantum-mechanical replacement (\ref{qmr}).
In particular, the energy density is 
\begin{subequations}
\be
T^{00} =\frac12 (\partial^0\phi)^2 +\frac12 (\nabla\phi)^2 +\frac12 V\phi^2
-\eta \nabla^2\phi^2.\label{sft00}
\ee
Equation (\ref{sft00}) is the form obtained directly by variation
of the Lagrangian with respect to  $g_{00}$,
but it can be rewritten using the equation of motion
(\ref{eom}), including the potential (\ref{fspot}),
without changing the numerical values of $T^{00}$.
For example, the equation of motion can be used to eliminate $V$ entirely:
\be
T^{00}=\frac12(\partial^0\phi)^2-\frac12\phi(\partial^0)^2\phi
+\frac14(1-4\eta)\nabla^2\phi^2,\label{t00}
\ee
which generalizes the flat-space analog of Eq.~(\ref{st00}).
If we take the vacuum expectation value of this, use the transform  
(\ref{ftt}), and integrate over all space, we immediately obtain,
for the spatial regulator, the energy
(\ref{regene}).

The form (\ref{t00}) does not mean that the energy density is free of 
 $\delta$ functions, however; from Eq.~(\ref{eom}) it is clear that the 
singularities in $V$ must be compensated by singularities in the
 second-order $z$ derivatives of $\phi$, hence
 ultimately of the reduced Green function~$g$.
 Note that such terms are absent from the part of
 Eq.~(\ref{t00}) that survives when $4\eta=1$.
 A case can  be made for using the equation of
 motion in the reverse direction in the remaining term,
replacing it as 
\be
T^{00}=\frac12(\partial^0\phi)^2-\frac12\phi(\partial^0)^2\phi
+\frac{1-4\eta}2\left[(\bm{\nabla}\phi)^2+ 
\phi(\partial_0)^2\phi+V\phi^2\right].\label{replace}
\ee
\end{subequations}
In this form the energy density that resides exactly
on the surface is exhibited explicitly by the $\delta$
functions in $V$, because (as will be verified)
the time derivatives and first-order space derivatives
are benign.  In particular, the surface energy arises
only when $4\eta\ne 1$.

To forestall confusion we must belabor two elementary
distinctions.  First, in the remainder of this section
we will see that the largest parts of the bulk energy  are concentrated
close to  the plates; such terms have  also  sometimes been called
``surface energy,''  but here we will reserve that term for energy
density that resides exactly on the surface.
 Second, because $\nabla^2\phi^2$
is a divergence, its integral over the region between
the plates (or the region to either side) can be
reduced to a surface integral over the plates;
but that is merely a mathematical representation of energy
that physically resides in the bulk.  However, as we will see
in Sec.~\ref{sec5}, this surface integral is another way of
describing the surface energy that resides on the plates.

\subsection{Energy density in bulk}

For purposes of calculation, we may use any of the forms of
the energy density given above, Eqs.~(\ref{sft00}), (\ref{t00}), or 
(\ref{replace}),
which directly lead to the following alternative
expressions for the  energy density in  ``reduced form'':
\begin{subequations}
\bea
t^{00}(z,z)&=&\frac12(-\zeta^2+k_\perp^2+\partial_z\partial_{z'})g(z,z')
\bigg|_{z'\to z}-\eta \partial_z^2g(z,z)+\frac12 V(z)g(z,z)\\
&=&-\zeta^2g(z,z)+\frac12(1-4\eta)(\partial_z^2+\partial_z\partial_{z'})g(z,z')
\bigg|_{z'\to z}\\
&=&-\zeta^2 g(z,z)+\frac12(1-4\eta)(\kappa^2+\partial_z\partial_{z'}+V)g(z,z')
\bigg|_{z'\to z}.
\eea
\end{subequations}
Deferring close examination of the surface terms to
Sec.~\ref{sec5}, we now study the energy density in the
regions excluding the plates themselves ($z\ne 0,\,a$):
\be 
u(z)=\langle T^{00}\rangle=u_{\rm int}(z)+u_1(z)+u_2(z),\label{localed}
\ee
where, excluding the Weyl term,
\begin{subequations}\label{leden}
\bea
u_{\rm int}(z<0)&=&\frac{\eta-1/6}{\pi^2}\int_0^\infty d\kappa\,\kappa^3
\lambda_2\left[\frac1{2\kappa+\lambda_1}\frac1\Delta
-\frac{e^{-2\kappa a}}{2\kappa+\lambda_2}
\right] e^{2\kappa z},\\
u_{\rm int}(0<z<a)&=&\frac1{\pi^2}\int_0^\infty d\kappa\,\kappa^3
\frac{\lambda_1\lambda_2}{(2\kappa)^2}\frac1\Delta
\left\{-\frac16+\left(\eta-\frac16\right)
\left[\frac1{1+2\kappa/\lambda_1}e^{-2\kappa z}
+\frac1{1+2\kappa/\lambda_2}
e^{-2\kappa(a- z)}\right]\right\},\\
u_{\rm int}(z>a)&=&\frac{\eta-1/6}{\pi^2}\int_0^\infty d\kappa\,\kappa^3
\lambda_1\left[\frac1{2\kappa+\lambda_2}\frac1\Delta
-\frac{e^{-2\kappa a}}{2\kappa+\lambda_1}
\right] e^{2\kappa (a-z)},
\eea
\end{subequations}
while the parts referring to each plate separately are
\be
u_1(z)=-\frac1{16\pi^3}\int_0^\infty d\kappa\,\kappa^3
\left[(1-4\eta)w(\pi/2)-f(\pi/2)\right]
\frac1{1+2\kappa/\lambda_1}e^{-2\kappa |z|},\label{divse}
\ee
while $u_2(z)$ is obtained from $u_1(z)$ by replacing $\lambda_1$
by $\lambda_2$ and $z$ by $a-z$.   (This is a symmetry of the energy density.)
In these equations $\Delta$ is still given by Eq.~(\ref{Delta})
and the  cutoff functions $w$ and $f$ are given in Eq.~(\ref{cutoffs}), 
When we are not too close to the plates, 
we can replace the cutoff function as follows,
\be
(1-4\eta)w(\pi/2)-f(\pi/2)\to-16\pi(\eta-1/6).
\ee
The replacement is valid for $|z|\gg\delta$ or $|z-a|\gg\delta$, 
where the integral over $\kappa$ are absolutely convergent.

\subsection{Strong coupling}
The integrals over $\kappa$
can be carried out explicitly in the case of strong coupling,
$\lambda_{1,2}\to\infty$.  In this Dirichlet limit, $u_2$ and $u_{\rm int}$
cancel for $z<0$, and only $u_1$ contributes there, while for $z>a$ only
$u_2$ survives.
(We will also see in Sec.~\ref{sec5} that the surface energy vanishes in strong
coupling.)
For example,
in strong coupling, the energy density below the plate is everywhere for a
spatial cutoff ($\gamma=\pi/2$)
\be
u(z<0)=\frac3{8\pi^2}\frac{(\eta-1/6)}{(z^2+\delta^2/4)^2}
+\frac1{32\pi^2}(1-4\eta)
\frac{\delta^2}{(z^2+\delta^2/4)^3},\label{ubelow1a}
\ee
which is finite as $z\to0$, and reduces to the familiar result 
\be
u(z<0)=\frac3{8\pi^2}\frac{\eta-1/6}{z^4},\label{ubelow2}
\ee
 if $|z|\gg\delta$.
Equation (\ref{ubelow1a}) agrees with the result given in Ref.~\cite{Estrada:2012yn} found for
a single plate for the special case $\eta=1/4$.  Above the top
plate, the energy density is given by the same expression (\ref{ubelow1a})
with $z\to a-z$. And in between, one finds
\be
u(0<z<a)=-\frac{\pi^2}{1440 a^4}+\frac{3(\eta-1/6)}{8\pi^2 a^4}\left[
\zeta(4,1+z/a)+\zeta(4,2-z/a)\right]+u(z<0)+u(z>a),
\ee
where the last two terms mean that the divergent terms (as $\delta\to0$)
are the same on both sides of the plates.
Here we have used the definition of the Hurwitz zeta function,
\be
\zeta(s,x)=\sum_{n=0}^\infty \frac1{(n+x)^s},\ee
which has the property
\be
\zeta(s,x)=\frac1{x^s}+\zeta(s,x+1).
\ee

Now when we integrate over all space, the Hurwitz
zeta functions telescope, and  we obtain the energy per unit area
\bea
\mathcal{E}-\mathcal{E}_W=-\frac{\pi^2}{1440 a^3}+\frac{\eta-1/6}{4\pi^2 a^3}
+\left[\int_{-\infty}^\infty +\int_{-a}^a\right]dz\left\{\frac3{8\pi^2}
\frac{\eta-1/6}{(z^2+\delta^2/4)^2}+\frac{\delta^2}{32\pi^2}\frac{1-4\eta}
{(z^2+\delta^2/4)^3}\right\}.\label{regD}
\eea
The integrals occuring here, for $\delta/a\to 0$, are
\be
\int_{-a}^a dz \frac1{(z^2+\delta^2/4)^2}=\frac{4\pi}{\delta^3}-\frac{2}{3a^3},
\quad
\int_{-a}^a dz\frac1{(z^2+\delta^2/4)^3}=\frac{12\pi}{\delta^5}-\frac2{5a^5}.
\ee
Thus the terms in Eq.~(\ref{regD})
proportional to $\eta-1/6$ cancel, including the term coming
from $1-4\eta=1/3-4(\eta-1/6)$, and we are left with
\be
\mathcal{E}-\mathcal{E}_W
=-\frac{\pi^2}{1440 a^3}+\frac1{4\pi\delta^3}\label{sce},
\ee  This gives us the Casimir interaction energy plus the self-energy of
both plates, twice Eq.~(\ref{sed}).

If we do the  temporal splitting, $\gamma=0$, 
the surface divergences are slightly modified.  Thus, for example,
\be
u(z<0)=\frac{3}{8\pi^2}\frac{\eta-1/6}{(z^2+\delta^2/4)^2}-\frac{\delta^2}{
8\pi}\frac\eta{(z^2+\delta^2/4)^3}.
\ee
So now when we integrate the energy density over all three regions the
$(1-4\eta)$ term instead has the factor $-4\eta$, so the self-energy term
in Eq.~(\ref{sce}) changes to 
\be
\mathcal{E}_{S_1+S_2}=-\frac1{2\pi\delta^3}.
\ee
This is exactly what is required by the recipe (\ref{0topi}) 
for passing from space-splitting to time-splitting, since the total
energy only depends on the cutoff function $f$.  Starting from the local
energy density, the total energy would therefore seem to  be given by
the regulated version of Eq.~(\ref{sec}), namely
\be
\mathcal{E}(\gamma)
=-\frac1{(2\pi)^3}\int_0^\infty d\kappa\,\kappa^4 f(\gamma)\Tr g,
\label{regenlocal}
\ee
which is different for $\gamma=0$ from that given in Eq.~(\ref{regene}).
Consistency, the perhaps dubious requirement that the regulated energy
have the same form,   suggests, therefore, that the spatial splitting 
$\gamma=\pi/2$
is preferred, the point being that calculating the energy from the
energy density leads to Eq.~(\ref{sec}), not the more stable Eq.~(\ref{fund}).

\subsection{Numerical results}
\begin{figure}
\centering
\includegraphics{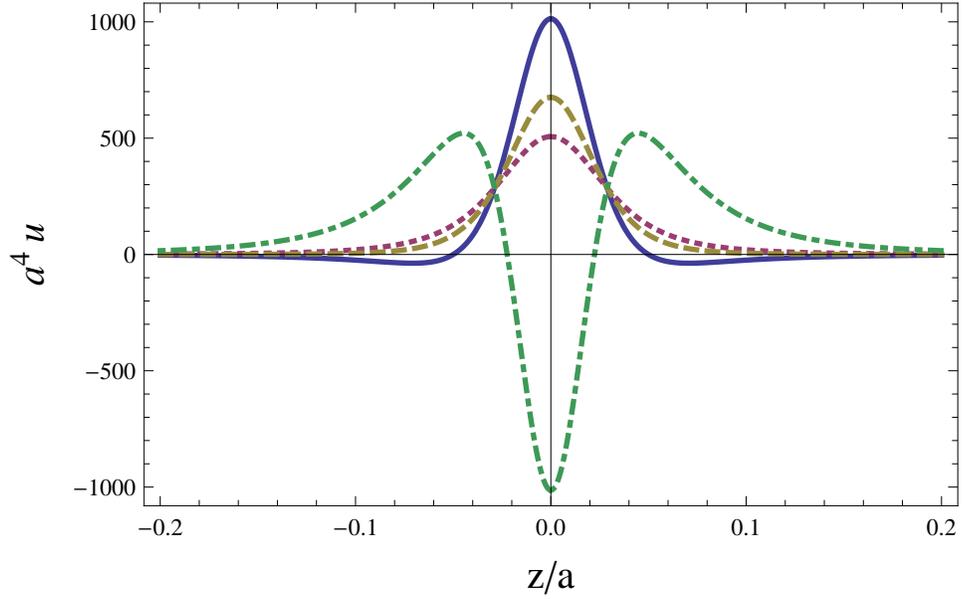}
\caption{\label{surfed} The energy density (in units of $a^4$)
near one of the Dirichlet plates.
Note that the 
energy density is very small except close to the plate.  The curves
are all for the cutoff $\delta=0.1 a$.  The curves are all for the spatial
splitting regularization, but for different values of the conformal parameter
$\eta$.  The solid (blue) curve is for the canonical case, $\eta=0$; the
short-dashed (red) curve is for $\eta=1/4$, where the surface energy is zero; 
the long-dashed (yellow) curve is for
$\eta=1/6$, the conformal value; and the dot-dashed, green curve is for 
$\eta=1$.  The energy per unit area
corresponding to any of these energy densities is
the same, $\mathcal{E}=79.5/a^3$. This is the value given by %one-half of
Eq.~(\ref{sed}), since the interaction energy density is negligible compared
to the self-energy densities.}
\end{figure}

The energy density for the Dirichlet limit and with $spatial$ splitting is
shown in Fig.~\ref{surfed}.
It is seen that in each case, the energy density is concentrated near the 
surfaces, and that when integrated, the rather different energy densities
correspond to a unique  energy per unit area  equal to that given by 
Eq.~(\ref{sce}).  In comparison, the interaction energy density is negligible.
This makes precise what we mean by saying that the
surface divergences are without consequence, giving rise
 to a self-energy of each plate, which can be considered as renormalizing
the mass of the plates.  Note that these self-energy densities do not vanish
when $\eta=1/6$, a fact that is completely overlooked by a naive
calculation without cutoff [see Eq.~(\ref{ubelow2})].

\begin{figure}
\centering
\includegraphics{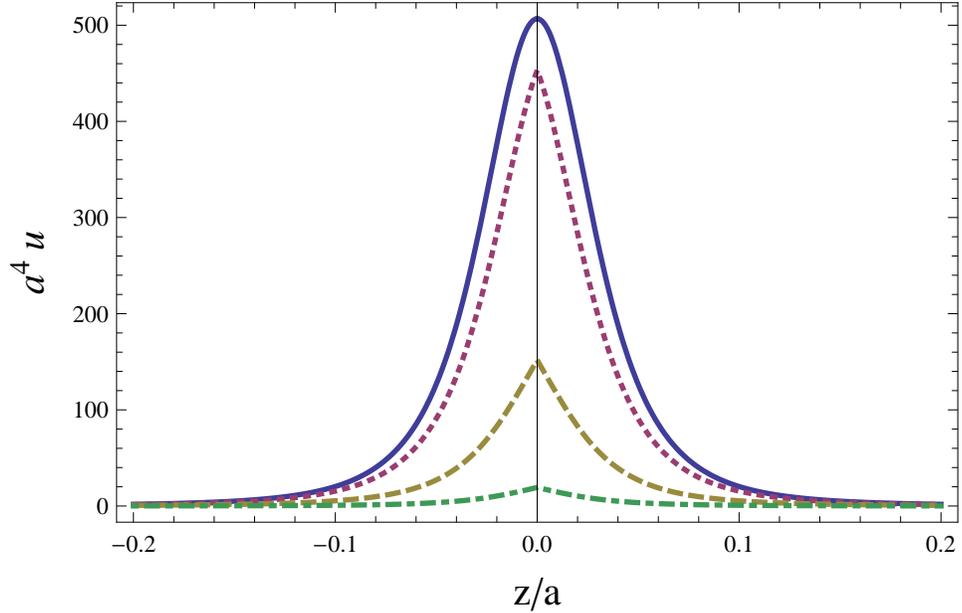}
\caption{\label{towardsc} The energy density (in units of $a^4$)
near one of the  plates for various values of the coupling $\lambda$.
Again the energy density is very small except close to the plate.  The curves
are all for the cutoff $\delta=0.1 a$.  They are all for the case where
the surface term is zero, $\eta=1/4$.   The solid (blue) curve is for 
the Dirichlet limit, $\lambda\to \infty$. The
short-dashed (red) curve is for $\lambda=100$; the long-dashed (yellow)
 curve is for $\lambda=10$,  and the dot-dashed (green) curve is for 
$\lambda=1$.  In all cases,  the interaction energy density 
is negligible compared to the self-energy density.
Consequently, the integrated energy in each case agrees with that
found from Eq.~(\ref{sefinl}).}
\end{figure}

For finite coupling we must proceed numerically.
In Fig.~\ref{towardsc} we similarly plot the Casimir energy density
for finite $\lambda$ for the
case when there is no surface term (as we shall see in the next section), that
is, when $\eta=1/4$ so the $\nabla^2\phi^2$ term in Eq.~(\ref{t00}) vanishes.
The energy density localized near the surfaces, corresponding to the cutoff-%
dependent terms $u_1$ and $u_2$, Eq.~(\ref{divse}), 
vastly dominate over the interaction energy
density.  As the coupling $\lambda\to\infty$, the Dirichlet limiting form
is rapidly approached.
 
\section{Surface terms}
\label{sec5}

In the previous section we only considered  points not on the plates at
$z=0$ and $z=a$.  But there are surface terms,
residing exactly on the plates, that need to be included to get the total
energy.
If we naively only included the integrated local energy density
in each region, and just dropped the divergent terms, we would get
 \be \int_{-\infty}^\infty  dz\,u(z)=
-\frac1{96\pi^3 a^3}\int_0^\infty dy\,y^3
\frac{1+\frac{12(\eta-1/6)}{y+\lambda_1 a}
+\frac{12(\eta-1/6)}{y+\lambda_2 a}}{\left(\frac{y}{\lambda_1 a}+1\right)\left(
\frac{y}{\lambda_2 a}+1\right)e^y-1},\ee
which disagrees with the correct interaction energy contained in 
Eq.~(\ref{regen}) except for $\eta=1/4$ or for $\lambda\to\infty$, 
where in either case the surface contribution to the energy vanishes.

The local surface energy density 
can be most easily found by using the energy density
in the form (\ref{replace}).
It is only the potential term (which of course vanishes off the plates)
that gives the surface energy: That due to the lower plate is therefore
\be
\Delta u_1(z)=\frac{1-4\eta}{2}\lambda_1\delta(z)\int_{-\infty}^\infty
\frac{d\zeta}{2\pi}\int\frac{(d\mathbf{k}_\perp)}{(2\pi)^2}
e^{i\bm{\kappa\cdot\delta}}g(0,0),
\ee
where we can take $g(0,0)$ to be given by Eq.~(\ref{geebelow}) in the limit as
$z=z'\to0$, since the Green's function is continuous.  This gives immediately
\be
\Delta u_1(z)=\delta(z)\frac{(1-4\eta)}{32\pi^3}\int_0^\infty d\kappa
\,\kappa\, w(\gamma)\frac{\lambda_1}{1+\lambda_1/2\kappa}
\left[1-\frac{\lambda_2}{2\kappa}\frac1\Delta\right],
\ee
and the energy density residing on the upper plate,  $\Delta u_2$, is given 
by a similar expression obtained by interchanging
$\lambda_1$ and $\lambda_2$ and replacing $z$ by $a-z$.  
Note that the first term in $\Delta u_i$ is a contribution to the self-energy,
while the second term contributes to the interaction energy.
Then the total energy density is, rather than that given in 
Eq.~(\ref{localed}),
\be
u(z)=u_{\rm int}+u_1(z)+\Delta u_1(z)+u_2(z)+\Delta u_2(z),
\ee
where $u_{\rm int}$ is given by Eq.~(\ref{leden}), and $u_1(z)$ by
Eq.~(\ref{divse}).
Integrating, we straightforwardly recover the total energy:
\bea
\int_{-\infty}^\infty dz \,u(z)
&=&\mathcal{E}-\mathcal{E}_W.\label{surfterm}\eea
This is exactly the result (\ref{regen}) obtained directly.

It appears that there is a self-energy 
contribution to the surface energy in the strong-coupling (Dirichlet) limit,
\be
\lambda\to\infty:\quad \Delta u_1(z)=\delta(z)\frac{1-4\eta}{16\pi^3}
\int_0^\infty d\kappa\,\kappa^2 w(\gamma).\ee
However,  using the expression for the cutoff function $w(0)=w(\pi/2)$ given in
Eq.~(\ref{transco}), we see that the integral here is zero:
\be
\int_0^\infty d\kappa\,\kappa^2 w(\pi/2)=4\pi\frac1\delta\frac{d}{d\delta}
\int_0^\infty d\kappa\cos\kappa\delta=0,
\ee
since the last integral vanishes in a distributional 
sense---See Ref.~\cite{Milton:2011iy}.
It is familiar that there should be no surface term for Dirichlet boundaries.

There is another, equivalent approach to the surface energy, which is
applicable to surfaces that are not described by potentials, such as
Robin boundaries.  It is known that, 
except in the Dirichlet (or Neumann) limit, one must include a term
 that resides exactly on the boundary 
\cite{Dowker:1978md,Kennedy:1979ar,Fulling:2003zx,Milton:2004vy,%
lebedev,romeo,saharian,bondurant}. 
This comes simply from integrating Eq.~(\ref{t00}) over some arbitrary
volume $V$ with boundary $\partial V$,
\be
\int_V(d\mathbf{r})\langle T^{00}\rangle=-\int_V(d\mathbf{r})\int_{-\infty}
^\infty \frac{d\zeta}{2\pi}\zeta^2\mathcal{G}(\mathbf{r,r})
+\frac{1-4\eta}{2}\int_{\partial V} d{\bf S}\cdot\bnabla 
\int_{-\infty}^\infty \frac{d\zeta}{2\pi}\mathcal{G}
(\mathbf{r,r'})\bigg|_{\mathbf{r'\to r}}.\label{genst}
\ee
The first term on the right is the total energy; the last term is the
negative of the
boundary energy.  [If we were to integrate over all space, including the
plates, the interior surface terms would disappear, and we would recover
the result (\ref{surfterm}).]
We can now apply this identity as follows.
Let the volume integral over the energy
density be only over the three regions outside the potentials, that is for
$z<0$, $0<z<a$, and $a<z$.  The surfaces at $z=0$ and $z=a$ are outside 
the region of the volume integration. 
Thus $\partial V$ are surfaces just above and 
below the $z=0$ and $z=a$ planes.
Because the Green's function in continuous, the total energy term is 
insensitive to the surfaces, which have measure zero.  On the other hand,
the boundary terms do not cancel, because the first derivatives are
discontinuous, so they give an additional contribution to the energy. 
If we call the boundary term $-\Delta\mathcal{E}$, we have
\be
\mathcal{E}=\int_V(d\mathbf{r})\langle T^{00}\rangle+\Delta \mathcal{E}.
\ee
The integral over the energy density in the bulk (i.e., excluding the plates)
must be supplemented by
the surface energy $\Delta\mathcal{E}$.
Combining the contributions coming from above and below
the two surfaces, we get here an additional
contribution to the energy that resides exactly on the surface:
\bea
\Delta \mathcal{E}&=&-\frac{1-4\eta}2\int_{-\infty}^\infty \frac{d\zeta}{2\pi}
\frac{(d\mathbf{k}_\perp)}{(2\pi)^2}e^{i\bm{\kappa\cdot\delta}}
\sum_{\rm plates} \mathbf{n}\cdot\bm{\nabla}g(z,z')\bigg|_{z'\to z}\nn\\
&=&-\frac{1-4\eta}{32\pi^3}\int_0^\infty
 d\kappa \,\kappa\, w(\gamma)\bigg[
-\frac{\lambda_1}{1+\lambda_1/2\kappa}-\frac{\lambda_2}{1+\lambda_2/2\kappa}
+\frac1\Delta\frac{\lambda_1\lambda_2}{2\kappa}\left(\frac1{1+
\lambda_1/2\kappa}+\frac1{1+\lambda_2/2\kappa}\right)\bigg].\label{surfterm0}
\eea
The sum  over $\mathbf{n}\cdot\bm{\nabla}$ on each plate signifies the
outward normal gradients
 from each region, $\mathbf{n}\cdot\bm{\nabla}=\pm\partial/
\partial z$, with the $+$ sign corresponding
to the boundary of the $z<0$ region at $z=0$, the $+$ and
$-$ signs referring to the boundaries of the $0<z<a$ region at $z=a$ and $z=0$,
respectively, and $-$ sign for the boundary of the $z>a$ region at $z=a$.
Note that the surface term depends only on the regulator function $w$ and
not on $f$, so it has the same value for both temporal and spatial splitting.
Not surprisingly, this agrees with our previous calculation,
\be
\Delta \mathcal{E}=\int_{-\infty}^\infty dz(\Delta u_1+\Delta u_2).
\ee

\section{How does surface energy fall?}\label{secfall}

Now we see that the arguments sketched in the introduction continue to
hold.  Either by looking in flat (Minkowski) space at the interaction
of the Casimir apparatus with a weak 
(Newtonian) gravitational field, or by working 
in Rindler coordinates and looking at the limit of small acceleration,
we see that the integral of the local energy density occurs, which gives
the total energy.  There are divergences in the local energy density
as the surfaces are approached, and
there are divergent contributions to the surface energy that live
entirely on the plates of the Casimir apparatus.  But we have regulated
the integrals with spatial and temporal cutoffs, and obtained
therefore  unique finite
values for the total energy. (The local energy density depends on the
conformal parameter.)
 The terms divergent as the cutoff goes to zero are contained in self-energies
serving to renormalize the masses of the plates, so are unobservable.  Both the
finite, cutoff-independent, Casimir interaction energy, and the divergent,
cutoff-dependent, self-energies, gravitate normally, that is, they obey the
equivalence principle.

To reiterate,  we have found an extremely simple answer to the question of how
Casimir energy gravitates: just like any other form of energy,
\be
\mathcal{F}=-g\mathcal{E}_C.
\ee
This result is independent of the orientation of the Casimir apparatus
relative to the gravitational field.  This refutes the claim sometimes
attributed to Feynman that virtual photons do not gravitate.
After a period of confusion, other authors agree with our conclusion
\cite{Bimonte:2006dv}.
However,  the previous arguments were formal, 
in that divergent self-energies were
not properly defined.  We have now regulated everything consistently, for 
both the global and local descriptions.  We have also considered arbitrary
conformal coupling parameter for the scalar field.
These calculations show,
quite generally, that the total Casimir energy, including the  divergent
parts, which  renormalize the masses of the plates,
possesses the gravitational mass demanded by the equivalence principle.
Similar conclusions were drawn by Saharian et al.\ \cite{Saharian:2003fd}
for the finite interactions between Dirichlet, Neumann, and conducting plates.
What is new in the present work is the explicit recognition that there
is a surface energy density residing on the Casimir plates,
which has been well defined through point-splitting regularization.  
When that term is included, the integrated energy
density equals the  total energy.  
Of course, if we considered only smooth
potentials, the surface energy would become continuously distributed throughout
the region of the potential.

\acknowledgments
KAM thanks the Laboratoire Kastler Brossel for their hospitality, particularly
Astrid Lambrecht and Serge Reynaud.  CNRS is thanked for their support.
This work was further supported in part by  grants from the 
U.S. National Science Foundation, the Simons Foundation,
and the Julian Schwinger Foundation;
earlier work summarized was supported by grants from
 the US Department of Energy.  We thank August Romeo  
and Jef Wagner for earlier collaborations
on this project and Hamilton Carter for comments on the manuscript.


\begin{thebibliography}{99}

\bibitem{casimir} H. B. G. Casimir, 
Kon.\ Ned.\ Akad.\ Wetensch.\ Proc.\  {\bf 51}, 793 (1948).

\bibitem{nernst} W. Nernst, Ver.\  Deut.\ Phys.\ Gesell.\ {\bf 18}, 83 (1921).

\bibitem{pauli} W. Pauli, {\it Handbuch der Physik\/} {\bf 24}, 83 (Springer,
Berlin, 1933). 

\bibitem{kragh} H. Kragh, Arch.\ Hist.\ Ex.\ Sci.\ {\bf 66}, 199 (2012)
[arXiv:1111.4623].




%\cite{Deutsch:1978sc}
\bibitem{Deutsch:1978sc} 
  D.~Deutsch and P.~Candelas,
  %``Boundary Effects in Quantum Field Theory,''
  Phys.\ Rev.\ D {\bf 20}, 3063 (1979).
  %%CITATION = PHRVA,D20,3063;%%
  %281 citations counted in INSPIRE as of 13 Nov 2013

%\cite{Milton:2004vy}
\bibitem{Milton:2004vy} 
  K.~A.~Milton,
  %``Casimir energies and pressures for delta function potentials,''
  J.\ Phys.\ A {\bf 37}, 6391 (2004)
  [hep-th/0401090].
  %%CITATION = HEP-TH/0401090;%%
  %38 citations counted in INSPIRE as of 19 Dec 2013

%\cite{Milton:2011iy}
\bibitem{Milton:2011iy} 
  K.~A.~Milton,
  %``Hard and soft walls,''
  Phys.\ Rev.\ D {\bf 84}, 065028 (2011)
  [arXiv:1107.4589 [hep-th]].
  %%CITATION = ARXIV:1107.4589;%%
  %3 citations counted in INSPIRE as of 13 Nov 2013

%\cite{Boyer:1968uf}
\bibitem{Boyer:1968uf} 
  T.~H.~Boyer,
  %``Quantum electromagnetic zero point energy of a conducting spherical shell and the Casimir model for a charged particle,''
  Phys.\ Rev.\  {\bf 174}, 1764 (1968).
  %%CITATION = PHRVA,174,1764;%%
  %298 citations counted in INSPIRE as of 13 Nov 2013



%\cite{Graham:2002fw}
\bibitem{Graham:2002fw} 
  N.~Graham, R.~L.~Jaffe, V.~Khemani, M.~Quandt, M.~Scandurra and H.~Weigel,
  %``Casimir energies in light of quantum field theory,''
  Phys.\ Lett.\ B {\bf 572}, 196 (2003)
  [hep-th/0207205].
  %%CITATION = HEP-TH/0207205;%%
  %74 citations counted in INSPIRE as of 13 Nov 2013

%\cite{Graham:2003ib}
\bibitem{Graham:2003ib} 
  N.~Graham, R.~L.~Jaffe, V.~Khemani, M.~Quandt, O.~Schroeder and H.~Weigel,
  %``The Dirichlet Casimir problem,''
  Nucl.\ Phys.\ B {\bf 677}, 379 (2004)
  [hep-th/0309130].
  %%CITATION = HEP-TH/0309130;%%
  %106 citations counted in INSPIRE as of 13 Nov 2013

\bibitem{barton}
G. Barton, J. Phys.\ A: Math.\ Gen.\ {\bf 37}, 1011 (2004).

\bibitem{barton2}
G. Barton, J. Phys.\ A: Math.\ Gen.\ {\bf 37}, 3725 (2004).




%\cite{Milton:2013yqa}
\bibitem{Milton:2013yqa} 
K.~A.~Milton, F.~Kheirandish, P.~Parashar, E.~K.~Abalo, S.~A.~Fulling, 
J.~D.~Bouas, H.~Carter and K.~Kirsten,
  %``Investigations of the torque anomaly in an annular sector. I. Global calculations, scalar case,''
  Phys.\ Rev.\ D {\bf 88}, 025039 (2013)
  [arXiv:1306.0866 [hep-th]].
  %%CITATION = ARXIV:1306.0866;%%
  %3 citations counted in INSPIRE as of 13 Nov 2013

%\cite{Milton:2013xia}
\bibitem{Milton:2013xia} 
  K.~A.~Milton, P.~Parashar, E.~K.~Abalo, F.~Kheirandish and K.~Kirsten,
  %``Investigations of the torque anomaly in an annular sector. II. Global calculations, electromagnetic case,''
  Phys.\ Rev.\ D {\bf 88}, 045030 (2013)
  [arXiv:1307.2535 [hep-th]].
  %%CITATION = ARXIV:1307.2535;%%
  %1 citations counted in INSPIRE as of 13 Nov 2013

%\cite{Estrada:2008zza}
\bibitem{Estrada:2008zza} 
  R.~Estrada, S.~A.~Fulling, Z.~Liu, L.~Kaplan, K.~Kirsten and K.~A.~Milton,
  %``Vacuum stress-energy density and its gravitational implications,''
  J.\ Phys.\ A {\bf 41}, 164055 (2008).
  %%CITATION = JPHGB,A41,164055;%%
  %10 citations counted in INSPIRE as of 13 Nov 2013

%\cite{Estrada:2012yn}
\bibitem{Estrada:2012yn} 
  R.~Estrada, S.~A.~Fulling, and F.~D.~Mera,
  %``Surface Vacuum Energy in Cutoff Models: Pressure Anomaly and Distributional Gravitational Limit,''
  J.\ Phys.\ A {\bf 45}, 455402 (2012)
  [arXiv:1207.7013 [gr-qc]].
  %%CITATION = ARXIV:1207.7013;%%
  %2 citations counted in INSPIRE as of 13 Nov 2013



%\cite{Fulling:2007xa}
\bibitem{Fulling:2007xa} 
S.~A.~Fulling, K.~A.~Milton, P.~Parashar, A.~Romeo, K.~V.~Shajesh, 
and J.~Wagner,
  %``How Does Casimir Energy Fall?,''
  Phys.\ Rev.\ D {\bf 76}, 025004 (2007)
  [hep-th/0702091].
  %%CITATION = HEP-TH/0702091;%%
  %45 citations counted in INSPIRE as of 13 Nov 2013

%\cite{Milton:2007ar}
\bibitem{Milton:2007ar} 
  K.~A.~Milton, P.~Parashar, K.~V.~Shajesh and J.~Wagner,
  %``How does Casimir energy fall? II. Gravitational acceleration of quantum vacuum energy,''
  J.\ Phys.\ A {\bf 40}, 10935 (2007)
  [arXiv:0705.2611 [hep-th]].
  %%CITATION = ARXIV:0705.2611;%%
  %20 citations counted in INSPIRE as of 13 Nov 2013

%\cite{Shajesh:2007sc}
\bibitem{Shajesh:2007sc} 
  K.~V.~Shajesh, K.~A.~Milton, P.~Parashar and J.~A.~Wagner,
  %``How does Casimir energy fall? III. Inertial forces on vacuum energy,''
  J.\ Phys.\ A {\bf 41}, 164058 (2008)
  [arXiv:0711.1206 [hep-th]].
  %%CITATION = ARXIV:0711.1206;%%
  %9 citations counted in INSPIRE as of 13 Nov 2013

\bibitem{brown}
L. S. Brown and G. J. Maclay, Phys.\ Rev.\ {\bf 184}, 1272 (1969).

%\cite{Scandurra:2000qz}
\bibitem{Scandurra:2000qz} 
  M.~Scandurra,
  %``Vacuum energy of a massive scalar field in the presence of a semitransparent cylinder,''
  J.\ Phys.\ A {\bf 33}, 5707 (2000)
  [hep-th/0004051].
  %%CITATION = HEP-TH/0004051;%%
  %20 citations counted in INSPIRE as of 19 Dec 2013

%\cite{CaveroPelaez:2006rt}
\bibitem{CaveroPelaez:2006rt} 
  I.~Cavero-Pel\'aez, K.~A.~Milton and K.~Kirsten,
%``Local and Global Casimir Energies for a Semitransparent Cylindrical Shell,''
  J.\ Phys.\ A {\bf 40}, 3607 (2007)
  [hep-th/0607154].
  %%CITATION = HEP-TH/0607154;%%
  %18 citations counted in INSPIRE as of 13 Nov 2013

\bibitem{moller}
C. M\o ller, {\it Theory of Relativity} (Oxford University Press, Oxford,
1972).



\bibitem{jaekel}
M.-T. Jaekel and S. Reynaud, J. Phys. I {\bf 3}, 1093 (1993).

%\cite{}
\bibitem{ccj} 
  C.~G.~Callan, Jr., S.~R.~Coleman and R.~Jackiw,
  %``A New improved energy - momentum tensor,''
  Ann.\ Phys.\ (N.Y.) {\bf 59}, 42 (1970).
  %%CITATION = APNYA,59,42;%%
  %658 citations counted in INSPIRE as of 28 Dec 2013

%\cite{Fulling:2003zx}
\bibitem{Fulling:2003zx} 
  S.~A.~Fulling,
  %``Systematics of the relationship between vacuum energy calculations and heat kernel coefficients,''
  J.\ Phys.\ A {\bf 36}, 6857 (2003)
  [quant-ph/0302117].
  %%CITATION = QUANT-PH/0302117;%%
  %64 citations counted in INSPIRE as of 13 Nov 2013


%\cite{Bordag:1998vs}
\bibitem{Bordag:1998vs} 
  M.~Bordag, K.~Kirsten and D.~Vassilevich,
  %``On the ground state energy for a penetrable sphere and for a dielectric ball,''
  Phys.\ Rev.\ D {\bf 59}, 085011 (1999)
  [hep-th/9811015].
  %%CITATION = HEP-TH/9811015;%%
  %107 citations counted in INSPIRE as of 22 Nov 2013



%\cite{Dowker:1978md}
\bibitem{Dowker:1978md} 
  J.~S.~Dowker and G.~Kennedy,
  %``Finite Temperature and Boundary Effects in Static Space-Times,''
  J.\ Phys.\ A {\bf 11}, 895 (1978).
  %%CITATION = JPHGB,A11,895;%%
  %235 citations counted in INSPIRE as of 13 Nov 2013

%\cite{Kennedy:1979ar}
\bibitem{Kennedy:1979ar} 
  G.~Kennedy, R.~Critchley, and J.~S.~Dowker,
  %``Finite Temperature Field Theory with Boundaries: Stress Tensor and Surface Action Renormalization,''
  Ann.\ Phys.\ (N.Y.) {\bf 125}, 346 (1980).
  %%CITATION = APNYA,125,346;%%
  %156 citations counted in INSPIRE as of 13 Nov 2013

\bibitem{lebedev}
S. L. Lebedev, %``Casimir effect in the presence of an elastic boundary.''
Zh.\ Eksp.\ Teor.\ Fiz.\ {\bf 110}, 769 (1996) 
[English transl.: JETP {\bf 83}, 423 (1996)].

\bibitem{romeo}
A. Romeo and A. Saharian, J. Phys.\ A {\bf35}, 1297 (2002).

\bibitem{saharian}
A. Saharian, Phys.\ Rev.\ D {\bf69}, 085005 (2004).

\bibitem{bondurant}
J. D. Bondurant and S. A. Fulling, J. Phys.\ A: Math.\ Gen.\ {\bf 38}, 1505
(2005).

%\cite{Bimonte:2006dv}
\bibitem{Bimonte:2006dv} 
  G.~Bimonte, E.~Calloni, G.~Esposito and L.~Rosa,
  %``Energy-momentum tensor for a Casimir apparatus in a weak gravitational field,''
  Phys.\ Rev.\ D {\bf 74}, 085011 (2006)
  %[Erratum-ibid.\ D {\bf 75}, 049904 (2007)]
  %[Erratum-ibid.\ D {\bf 75}, 089901 (2007)]
  [Erratum-ibid.\ D {\bf 77}, 109903 (2008)]
  [hep-th/0606042].
  %%CITATION = HEP-TH/0606042;%%
  %24 citations counted in INSPIRE as of 13 Nov 2013

%\cite{Saharian:2003fd}
\bibitem{Saharian:2003fd} 
  A.~A.~Saharian, R.~S.~Davtyan and A.~H.~Yeranyan,
  %``Casimir energy in the Fulling-Rindler vacuum,''
  Phys.\ Rev.\ D {\bf 69}, 085002 (2004)
  [hep-th/0307163].
  %%CITATION = HEP-TH/0307163;%%
  %20 citations counted in INSPIRE as of 13 Nov 2013
y\end{thebibliography}
\end{document}